\documentclass[a4paper,10pt,twocolumn]{article}
\usepackage[utf8x]{inputenc}
\pdfoutput=1
\usepackage{amsmath,mathrsfs,latexsym,amssymb,graphicx,xcolor,float}
\usepackage[square,comma,numbers,sort&compress]{natbib}

\newcommand{\be}{\begin{equation}}
\newcommand{\ee}{\end{equation}}

\newcommand{\dl}{\mathrm{\delta}}
\newcommand{\oter}[2]{|#1 \rangle \langle #2|}

\newcommand{\kett}[1]{|#1 \rangle}

\newcommand{\uut}[2]{#1_{_#2}(t)}
\newcommand{\uu}[2]{#1_{_#2}}
\newcommand{\ud}{\mathrm{d}}
\newcommand{\avgf}[1]{\left\langle#1\right\rangle_{_{\!\!f}}}
\newcommand{\avguf}[1]{\left\langle#1\right\rangle_{_{\!\!\!\!f}}}
\newcommand{\ff}{{f_0^2}}
\graphicspath{{figs/}}

\begin{document}

\title{Effect of Injected Noise on \\ Electromagnetically Induced Transparency
and Slow Light}
\author{V. Ranjith\footnote{For correspondence~: ranjithv AT rri.res.in} \ and N. Kumar \\
Raman Research Institute, Bangalore 560 094, India. \\
}
\date{}
%\section**{}
\maketitle

\begin{abstract}
We have examined theoretically the phenomenon of Electromagnetically Induced Transparency (EIT) in a three--level system operating in the $\Lambda$--configuration %(with the level energies {$E_1<E_2<E_3$}) %
in the presence of an externally injected noise coupling the ground level %($\kett{1}$)
to the intermediate (metastable) level. %$\kett{2}$.
The changes in the depth and the width of the induced transparency and the slowing down of the probe light have been calculated as function of the probe detuning and the strength of the injected noise. The calculations are within the rotating-wave approximation (RWA). Our main results are the reduction and the broadening of the  EIT with increasing strength of the injected noise, and a reduction in the slowing down of group velocity of the probe-laser beam. Thus, the injected semi-classical noise, unlike the quantum-dynamical noise associated with the spontaneous emission, is not effectively cancelled by the EIT mechanism.

\end{abstract}

\section*{Keywords }
%Electromagnetically induced transparency, EIT, Rabi frequency, gaussian white noise, dispersion and slow light, density matrix, lindblads, Novikov theorem.
EIT, group velocity, Lindblad, \mbox{Novikov's theorem,} gaussian white noise.

\section*{Introduction}
Electromagnetically Induced Transparency (EIT) is a coherent quantum optical phenomenon in which the absorption of a weak probe beam of light vanishes, opening thereby a window of transparency narrower than the natural linewidth in the center of the  otherwise much broader resonance absorption peak$^{[1,2]}$. This transparency is due to the destructive quantum interference of the transition amplitudes along the allowed alternative paths, much as in the case of the well-known Fano resonance/anti-resonance$^{[3,4]}$. This phenomenon is known to give rise to several other interesting effects$^{[1,2,5]}$, e.g., the slow light. Remarkably, this interference {effectively cancels the spontaneous down-transitions} (quantum noise$^{[6]}$) as well, giving a sub-natural linewidth as noted above. In contrast to this, as the present work shows, {the EIT is indeed affected/degraded by injecting a  'classical' noise into one of the alternative paths.}

The purpose of this work is to analyze theoretically the nature and the magnitude of the effect of the injected noise on the depth and the width of the transparency (EIT), as well as the associated change in the group velocity of the probe light as function of the strength of noise and the probe detuning. {The calculations have been done using the density-matrix formalism wherein the various natural linewidths (spontaneous decays) involved are introduced via the  appropriately chosen Lindblad super-operators$^{[6]}$, while the injected noise $f(t)$ is introduced explicitly in the Hamiltonian.} The latter is treated within the Rotating Wave Approximation (RWA). We compute the reduced density-matrix averaged over the noise $f(t)$. This could be carried out here in a closed form by assuming the injected noise to be a Gaussian White Noise (GWN), and using the well-known Novikov theorem${[7]}$. The latter holds for averaging an arbitrary functional of the gaussian white noise. The physical quantities of interest are then calculated in terms of this noise-averaged density matrix. Our main results are a quantitative reduction and broadening of the EIT with increasing strength of the injected noise, and a reduction in the slowing down of the group velocity of the probe light.

\section*{The atomic-level scheme}
The atomic-level scheme considered here is as in a typical EIT experiment shown schematically in \mbox{Figure 1}. It involves a manifold of three energy levels $\{\kett{1},\kett{2},\kett{3}\}$ with the respective energies \mbox{$E_1=\hbar\omega_1$, $E_2=\hbar\omega_2$, and $E_3=\hbar\omega_3$}, connected in the $\Lambda$--configuration with \mbox{$E_1<E_2< E_3$}. Such a $\Lambda$--EIT scheme is possible, e.g., in the D2-line transitions of ${}^{85}\mathrm{Rb}$ vapor$^{[8,9]}$ with \mbox{$\kett{1} \equiv 5^2\mathrm{S_{1/2}:F=2}$,} \mbox{$\kett{2} \equiv 5^2\mathrm{S_{1/2}:F=3}$,} and \mbox{$\kett{3} \equiv 5^2\mathrm{P_{3/2}:F'=3}$}. As follows from the selection rules, in a $\Lambda$ scheme, the transition (here $\kett{1}\leftrightarrow\kett{2}$)  between the ground and the intermediate state is necessarily an electric--dipole forbidden transition, since the other two transitions involved are electric--dipole allowed. It can, however, be, e.g., an electric-quadrupole allowed transition, much as in the case of a controlling microwave field used in the recent EIT--experiment in rubidium vapor system by Hebin Li., et al.$^{[10]}$. A novel feature of the present work is a noise field $f(t)$ injected externally at the transition \mbox{$\kett{1}\leftrightarrow\kett{2}$}.
\begin{figure}[H]
\centering
 \includegraphics[width=.41\textwidth]{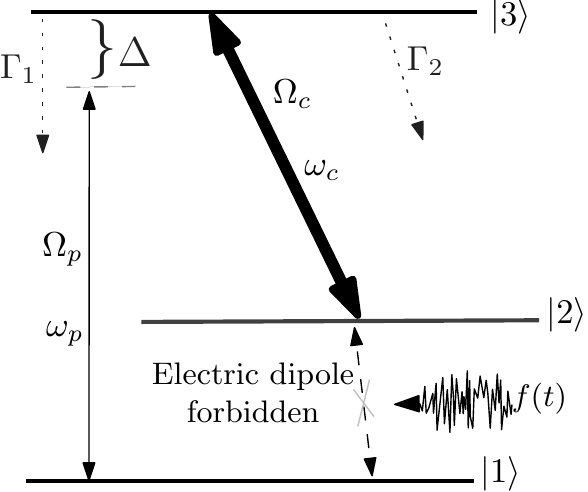}
\caption{\small EIT scheme of $\Lambda$-type three--level system perturbed by an externally injected noise $f(t)$.}
\end{figure}
In a standard EIT experiment, the absorption of the probe beam is studied as function of its detuning while the coupling laser is kept at resonance$^{[2]}$. Accordingly, here the probe laser is detuned off--resonance by an amount $\Delta$ from the transition \mbox{$\kett{1}\rightarrow \kett{3}$} which is being probed, i.e., \mbox{$\hbar \omega_p = E_3 - E_1 - \hbar \Delta$}, where $\omega_p$ and $\Omega_p$  are, respectively, the frequency and the associated Rabi-frequency of the probe laser. A strong coupling laser beam is applied and kept at resonance to the transition $\kett{2}\leftrightarrow\kett{3}$, with \mbox{$\hbar \omega_c = E_3- E_2$}, where $\omega_c$ and $\Omega_c$ are, respectively, the frequency and the associated Rabi frequency of the coupling laser. Further, $\Gamma_1$ and $\Gamma_2$ denote the rates of spontaneous decay of the excited level $\kett{3}$ to the ground level $\kett{1}$ and to the intermediate level $\kett{2}$ respectively.
%In addition to the coherent laser beams, a gaussian white noise $f(t)$ is injected into the transition $\kett{1}\leftrightarrow\kett{2}$.

\section*{The Hamiltonian and RWA}
%% FULL HAMILTONIAN, INTERACTION-PICTURE HAMILTONIAN, RWA-HAMILTONIAN %%

The full Hamiltonian ${{\mathscr{H}}}(t)$ of the EIT system here corresponds essentially to the case of a three--level system being acted upon by two optical fields in the so-called semi-classical approximation, but now with an extra feature, namely, a term $f(t)$ corresponding to the injected noise. More explicitly, we have
\be \begin{split} {\mathscr{H}}(t)~=&~\hbar \omega_1\oter{1}{1}+ \hbar \omega_2\oter{2}{2} + \hbar \omega_3\oter{3}{3} \\ 
& -\hbar \Omega_p (e^{-i\omega_p t}+e^{i\omega_p t})~\left(\oter{3}{1}+\oter{1}{3}\right)\\
&-\hbar \Omega_c (e^{-i\omega_c t}+e^{i\omega_c t})~\left(\oter{3}{2} + \oter{2}{3}\right)\\ 
& -\hbar f(t)~\left(\oter{1}{2}+\oter{2}{1}\right)~. \end{split} \ee
Let $${\mathscr{H}}(t)~\equiv~{\mathscr{H}}_{0} + {\mathscr{H}}_{1}(t)$$ \\
with 
\be {\mathscr{H}}_{0} =  -\hbar\omega_p \oter{1}{1} - \hbar \omega_c \oter{2}{2}~,\ee 
and \\
\be \begin{split} {\mathscr{H}}_{1}(t) =&~\hbar \left( \omega_1+\omega_p \right) \oter{1}{1}~+~\hbar\left(\omega_2+\omega_c\right) \oter{2}{2} \\
&+\hbar \omega_3~\oter{3}{3}~-~\hbar f(t)\left(\oter{1}{2}+\oter{2}{1}\right)\\ 
&-\hbar \Omega_p (e^{-i\omega_p t}+e^{i\omega_p t})~(\oter{3}{1}+\oter{1}{3})\\
&-\hbar \Omega_c (e^{-i\omega_c t}+e^{i\omega_c t})~(\oter{3}{2} + \oter{2}{3})~.\\ 
\end{split} \ee
We now proceed to the interaction(I) picture with the corresponding Hamiltonian $\mathscr{H}_{I}(t)$ given by
\be {\mathscr{H}}_{I}(t) = e^{i{\mathscr{H}}_{0}t/\hbar} {\mathscr{H}}_{1}(t) e^{-i{\mathscr{H}}_{0}t/\hbar}. \ee
In the Rotating Wave Approximation (RWA) (i.e., neglecting the rapidly oscillating terms \mbox{proportional} to $e^{\pm2i\omega_ct}~and~e^{\pm2i\omega_pt}$), we obtain
\be \begin{split} {\mathscr{H}}_{I}(t)=
&-\hbar f(t)\left(e^{-i\omega_{\mu} t} \oter{1}{2}+e^{i\omega_{\mu} t} \oter{2}{1}\right)\\
&-\hbar \Omega_p (\oter{3}{1}+\oter{1}{3}) - \hbar \Omega_c (\oter{3}{2} + \oter{2}{3})\\ 
&+\hbar \left( \omega_1+\omega_p \right) \oter{1}{1}~+~\hbar\left(\omega_2+\omega_c\right) \oter{2}{2} \\
&+\hbar \omega_3\oter{3}{3} \end{split} \ee
with $\omega_{\mu} \equiv \omega_2-\omega_1 -\Delta$. \\
Hereinafter, we will use the interaction-picture Hamiltonian ${\mathscr{H}}_{I}(t)$, but drop the subscript $I$ for convenience.

\section*{Lindblad EoMs}
%%Lindblad equation, EoMs before averaging.

%Hereinafter we will use the interaction-picture Hamiltonian ${\mathscr{H}}_{I}(t)$, dropping the subscript $I$ for convenience.

With this, the master equation of motion for the density matrix $\rho(t)$ is
\be \dot{\rho}(t) = -\frac{i}{\hbar}\left[{\mathscr{H}}(t),\rho(t)\right]- \widehat{\Lambda} \rho(t), \ee
where we have introduced the Linblad super-operator $\widehat{\Lambda}$, given in the diagonal form as
\be  \widehat{\Lambda} \rho \equiv \sum_{j=1,2} \Gamma_j \left(L_j^{\dagger}L_j\rho + \rho L_j^{\dagger}L_j -2 L_j \rho L_j^{\dagger}\right). \ee
Here $\Gamma_1$ and $\Gamma_2$ are the rates of spontaneous decays $\kett{3} \rightarrow \kett{1}$ and $\kett{3} \rightarrow \kett{2}$ respectively.
The Lindblad operators ${L}_1$ and ${L}_2$  are chosen appropriately so as to describe the spontaneous decays as$^{[6]}$:
\be L_1 = \oter{1}{3}, \ee
\be L_2 = \oter{2}{3}. \ee
As is known well, a Lindblad master equation describes the non-unitary evolution of the density matrix preserving the trace condition without violating its complete positivity and hermiticity  for all initial conditions${^{[6]}}$.

Substituting from eqs.(5,7) into eq.(6), we obtain
\be \begin{split} -i\,\dot{\rho}_{_{11}}(t) = 
&~f(t)\left( e^{-i\omega_{\mu} t} \rho_{_{21}}(t)-e^{i\omega_{\mu} t}\rho_{_{12}}(t)\right) \\
& + \Omega_p (\rho_{_{31}}(t)-\rho_{_{13}}(t)) -i \Gamma_{_1} \rho_{_{33}}(t), \end{split} \ee
\be \begin{split} -i\,\dot{\rho}_{_{22}}(t) = 
&~f(t)\left( e^{i\omega_{\mu} t} \rho_{_{12}}(t)-e^{-i\omega_{\mu} t}\rho_{_{21}}(t)\right) \\
& + \Omega_c (\rho_{_{32}}(t)-\rho_{_{23}}(t)) -i \Gamma_{_2} \rho_{_{33}}(t), \end{split} \ee
\be \begin{split} -i\,\dot{\rho}_{_{12}}(t) = 
&~\Delta~\rho_{_{12}}(t) +\Omega_p \uut{\rho}{{32}}-\Omega_c \uut{\rho}{{13}} \\
&+f(t) e^{-i\omega_{\mu} t} (\uut{\rho}{{22}}-\uut{\rho}{{11}}), \end{split} \ee

\be \begin{split} -i\,\dot{\rho}_{_{13}}(t) = 
&~f(t) e^{-i\omega_{\mu} t} \uut{\rho}{{23}} -\Omega_c \uut{\rho}{{12}} \\
&+\Omega_p(\uut{\rho}{{33}}-\uut{\rho}{{11}}) \\
&+ \left(\Delta +i\frac{\Gamma_{_1}+\Gamma_{_2}}{2}\right)\uut{\rho}{{13}},  \end{split} \ee
\be \begin{split} -i\,\dot{\rho}_{_{23}}(t) = 
&~i\frac{\Gamma_{_1}+\Gamma_{_2}}{2} \uut{\rho}{{23}}+f(t) e^{i\omega_{\mu} t} \uut{\rho}{{13}} \\
&+ \Omega_c\left(\uut{\rho}{{33}}-\uut{\rho}{{22}}\right)-\Omega_p \uut{\rho}{{12}}, \end{split} \ee
along with the trace condition $Tr\rho = 1$ and the hermiticity condition $\rho^\dagger = \rho$.

\section*{Averaging the EoMs over noise}
%Novikov theorem, noise averaged Density matrix EoMs.

%(i.e., $\rho(t) \equiv \rho[f(t)]$)

Now, the density matrix $\rho(t)$ is a functional of the noise $f(t)$ occurring in eqs. (10-14), and, therefore, it must be averaged over all the realizations of $f(t)$. For this, we have taken the noise $f(t)$ to be a Guassian White Noise (GWN), i.e.,
\be \avgf{ f(t) f(t') } = f_0^2\,\delta(t-t') ,\ee
where $f_0^2$, having the dimension of frequency, is a measure of the strength of the noise $f(t)$ (much like the Rabi frequency $\Omega$, which is a measure of the strength of the a laser field). For this, we make use of the Novikov theorem$^{[]}$ giving
\be \begin{split}
\avgf{ \, f(t) \, \rho[f(t)] \, } 
&~=\int_{_{\!-\infty}}^{t} \!\! \ud t'\, \avgf{ f(t) f(t') } \, \avguf{ \frac{\dl \rho[f(t')]}{\dl f(t)} } \\
&=~\frac{f_0^2}{2} \avguf{{\frac{\dl \rho[f(t)]}{\dl f(t)}}}.\end{split} \ee
 
Straightforward functional differentiation of eqs.(10-14) w.r.t. $f(t)$ and using eq.(16), we obtain, within RWA (this time neglecting the rapidly oscillating terms \mbox{proportional} to $e^{\pm2i\omega_\mu t}$), a closed set of equations for the noise-averaged density matrix elements $\avgf{\rho_{_{ij}}(t)}$ ($\equiv \sigma_{_{ij}}(t)$ for typographic convenience) as follows :
\be \begin{split} -i\,\dot{\sigma}_{_{11}}(t) = 
&~\Omega_p (\sigma_{_{31}}(t)-\sigma_{_{13}}(t))-i\Gamma_{_1}\sigma_{_{33}}(t)  \\
&-i\ff\left(\uut{\sigma}{{22}}-\uut{\sigma}{{11}}\right), \end{split} \ee
\be \begin{split} -i\,\dot{\sigma}_{_{22}}(t) = 
&~\Omega_c (\sigma_{_{32}}(t)-\sigma_{_{23}}(t))-i \Gamma_{_2} \sigma_{_{33}}(t)\\
&+i\ff\left(\uut{\sigma}{{22}}-\uut{\sigma}{{11}}\right), \end{split} \ee
\be \begin{split} -i\,\dot{\sigma}_{_{12}}(t)~=~ 
&~\left(\Delta + i\ff\right)\sigma_{_{12}}(t)~+~\Omega_p \uut{\sigma}{{32}}~\\
&-\Omega_c \uut{\sigma}{{13}}, \end{split} \ee
\be \begin{split} -i\,\dot{\sigma}_{_{13}}(t) = 
&-\Omega_c \uut{\sigma}{{12}} + \Omega_p(\uut{\sigma}{{33}}-\uut{\sigma}{{11}}) \\
&+\left(\Delta + i \frac{\ff+\Gamma_{_1}+\Gamma_{_2}}{2} \right) \uut{\sigma}{{13}}, \end{split} \ee
\be \begin{split} -i\,\dot{\sigma}_{_{23}}(t) = 
&~\Omega_c\left(\uut{\sigma}{{33}}-\uut{\sigma}{{22}}\right)-\Omega_p \uut{\sigma}{{12}} \\
&+i\left(\frac{\ff + \Gamma_{_1}+\Gamma_{_2}}{2} \right) \uut{\sigma}{{23}}. \end{split} \ee

These linear first-order differential equations, along with the trace condition $\mathrm{Tr}\sigma = 1$ and the hermiticity condition $\sigma^\dagger = \sigma$, are now solved numerically in the steady state \mbox{(i.e., $\dot{\sigma}(t)=0$)} so as to calculate the physical quantities of interest, as described below.

\section*{Results}
The absorption coefficient of the probe light in a dilute gaseous medium can be expressed in terms of the density matrix as $^{[5]}$
\be \alpha(\Delta,\Omega_c,\ff)= \frac{1}{\lambda_p} \frac{N \lambda^3_{0} \pi}{(\Omega_p/\Gamma)} \mathrm{Im}\left[\uu{\sigma}{{31}}(\Delta,\Omega_c,\ff)\right] \, ,\ee
where, $N$ is the atomic number density of the medium,  $\lambda_p$ is the wavelength of the probe light at a detuning $\Delta$, and $\lambda_0 = \frac{c}{( \omega_3 - \omega_1)}$ is the wavelength at the corresponding resonance. Also, we have set $\Gamma_2 = 0$ (which is in fact a good approximation for most of the practical $\Lambda$-type EIT systems), and $\Gamma_1 \equiv \Gamma$.
The corresponding real part of the refractive index can be expressed in terms of the various parameters involved$^{[5]}$ as 
\be \uu{n}{R} (\Delta,\Omega_c,\ff) = 1 + \frac{N \lambda^3_{0} \pi}{(\Omega_p/\Gamma)}  \mathrm{Re}\left[\uu{\sigma}{{31}}(\Delta,\Omega_c,\ff)\right] \, . \ee
The associated group velocity of the probe light in the medium concerned can be conveniently expressed as$^{[5]}$:
\be V_g(\Delta,\Omega_c,\ff) = \frac {c}{n_{R}(\Delta,\Omega_c,\ff) - \omega_p \frac{\ud \uu{n}{R}}{\ud \Delta}}. \ee

In the following Figures 2-6, we have plotted these various physical quantities of interest ($\alpha\,,~\uu{n}{R}\,\mathrm{~and~}V_g$), calculated as functions of the parameters involved (i.e., the probe-detuning$\,\Delta/\Gamma$, the coupling strength$\,\Omega_c/\Gamma$, and the strength of noise$\,\ff/\Gamma$). For this, we have used the following set of parameter values appropriate to the EIT-medium, ${}^{85}\mathrm{Rb}$ vapuor$^{[8,9]}$ : \\

\begin{tabular}{r c l} 
$N$				&= 	&$10^{18}$ atoms per $m^3$, \\
$\lambda_0$			&=	&$780\mathrm{nm,}$ \\
$({\Omega_c}/{\Gamma})$		&=	&$1,$ \\
$({\Omega_p}/{\Gamma})$		&= 	&$0.001,$ \\
$({\ff}/{\Gamma})$		&= 	&$\{ 0, 0.7, 1.6\},$ \\
$\Gamma$			&= 	&$5 \mathrm{MHz}$.\\
\end{tabular}

%>>>>>>>>>>>>>>>>>>>>>>>>>>>>
\begin{figure}[H]
\centering
 \includegraphics[width=.46\textwidth]{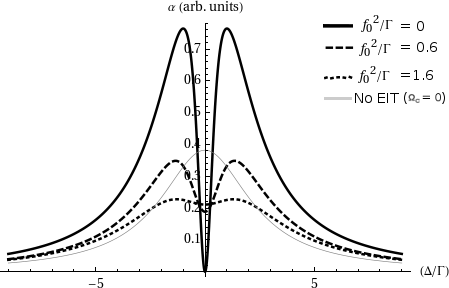}
\caption{\small Absorption(arb. units) of probe light plotted against the detuning $(\Delta/\Gamma)$ for different values of the strength of noise $(\ff/\Gamma)$.}
\end{figure}

\begin{figure}[H]
\centering
 \includegraphics[width=.46\textwidth]{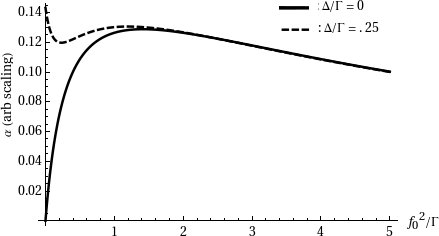}
\caption{\small Absorption(arb. units) of probe light at and near resonance, plotted against the strength of noise $(\ff/\Gamma)$.}
\end{figure}

%\textbf{Discussions:}

From Figure 2, it can be readily seen, as expected for pure EIT (without noise), that the absorption of the probe light beam increases with detuning ($\Delta/\Gamma$), for small values of the detuning. 

\noindent With the increasing strength of the noise ($\ff/\Gamma$), however,  the absorption as well as its spectral width increases. Overall, the effect of noise is more pronounced within the EIT window, and much less so outside the window, as clearly seen in Figure 3.

%>>>>>>>>>>>>>well as the

\begin{figure}[H]
\centering
 \includegraphics[width=.46\textwidth]{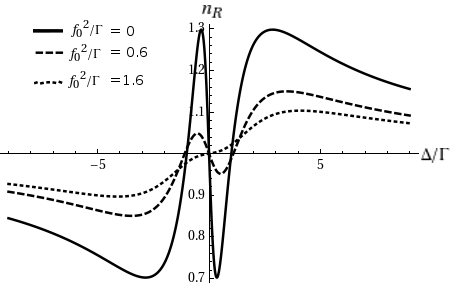}
\caption{\small Real part of the refractive index $(\uu{n}{R})$ plotted against the detuning $(\Delta/\Gamma)$, for different values of the strength of noise $(\ff/\Gamma)$.}
\end{figure}
%\textbf{Discussions :}

Figure 4 gives specifically the variation of the real part of the refractive index ($\uu{n}{R}$) as function of detuning in the anomalous regime of dispersion, within the EIT window. There is a pronounced decrease in the variation of $\uu{n}{R}$ across the EIT window with the increasing strength of noise.

%>>>>>>>>>>>>>>>>>>>>>>>
Now we come to the effect of noise on the group velocity of the probe light beam across the EIT window. This has an important bearing on the phenomenon of slow light associated with the EIT.
\begin{figure}[H]
\center
 \includegraphics[width=.46\textwidth]{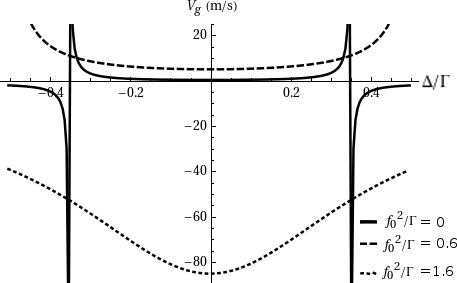}
\caption{\small Group velocity ($V_g$ m/s) of the probe light in the medium, plotted against the detuning $(\Delta/\Gamma)$, for different values of the strength of noise $(\ff/\Gamma)$.}
\end{figure}

\begin{figure}[H]
\centering
 \includegraphics[width=.46\textwidth]{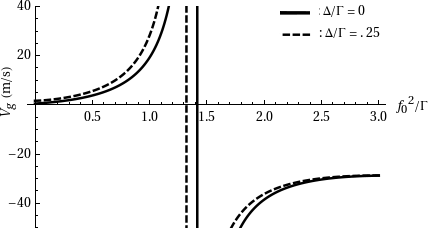}
\caption{\small Group velocity ($V_g$ m/s) of the probe light in the medium at and near resonance, plotted against the strength of noise $(\ff/\Gamma)$.}
\end{figure}
 The overall effect of the noise is to reduce the slowing down of light beam as can be seen from the Figure 5. Thus, e.g., $\mathrm{V_g=0.52 ms^{-1}}$~ for zero noise and zero detuning). Figure 6 resolves the finer features of the effect of the noise on the group velocity at and near the resonance (i.e., $\Delta/\Gamma=0, 0.25$).
%\textbf{Concluding remarks:}  

Finally, a few words are in order to explain how the injected noise enters the physics of the EIT. Basically, the injected noise is transfered by the coupling laser beam into the excited level $\kett{1}$, giving rise to two related effects : First, it reduces (dephases) the interference effects responsible for the EIT itself; and secondly, it broadens the excited energy level$\kett{1}$. These effects in turn reduce the depth of the EIT window while enhancing its width. This transfer effect is  seen to be reflected in Figure 7.

\begin{figure}[H]
\centering
 \includegraphics[width=.40\textwidth]{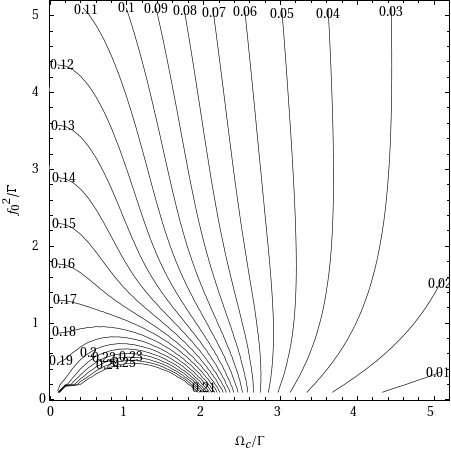}
\caption{\small Contour-plot of constant absorption coefficient ($\alpha$) in the $(\Omega_c/\Gamma)-(\ff/\Gamma)$ plane. The labels on contours indicate the $\alpha$ values (in arbitrary units)}
\end{figure}

\rule{.47\textwidth}{.1pt}%
\section*{Bibliography}
\begin{enumerate}
\item Harris, S., Electromagnetically induced transparency. \textit{Phys. Today}, 1997, 50, 36–42.

\item Fleischhauer M., Imamoglu A., and Marangos, J.P., Electromagnetically induced transparency : Optics in coherent media.
 \textit{Rev. Mod. Phys.}, 2005, 77, 633.

\item Fano, U., Effects of configuration interaction on intensities and phase shifts. \textit{Phys. Rev.}, 1961, 124, 1866–1878.

\item Cohen-Tannoudji, C., Dupont-Roc, J., Grynberg, G., ``Atom-photon interactions: basic processes and applications''. \textit{Wiley science--Paperback series}, Wiley,  1998.

\item Milonni, P.W., ``Fast light, Slow light and Left-handed light''. \textit{Series in Opt. and Optoelect.},  Taylor $\and$ Francis Group.(2005).

\item Gardiner, C.W. and Zoller, P., ``Quantum noise: a handbook of Markovian and non-Markovian quantum stochastic methods with applications to quantum optics''. 2004, 2nd edn., Springer.

\item E. A. Novikov.,  \textit{Zh. Eksp. Teor.Fiz.}, 1964, 47, 1919. ; [\textit{Sov. Phys. JETP.}, 1965, 20, 1990].

\item Abraham, J.O. and Mayer, S. K., Electromagnetically induced transparency in rubidium.  \textit{Am. J. Phys.}, 2009, 77, 116-121.

\item Steck D.A., ''Rubidium 85 D Line Data'' \textit{http://steck.us/alkalidata/rubidium85numbers.pdf}

\item Hebin Li., et al., Electromagnetically induced transparency controlled by a microwave field. \textit{Phys. Rev. A}, 2009, 80, 023820.
\end{enumerate}

%%%%%%%%%%%%%%%%%%%%%%%%%%%%%%%%%%%%%%%%%%%%%%%%%%%%%%%%%%%%%%%%%%%%%%%%%%%%%%%%%%%%%

\rule{.45\textwidth}{.1pt}%

\noindent \textbf{Acknowledgements:} We would like to thank \mbox{Andal Narayanan} for fruitful discussions. One of us (RV) thanks the Raman Research Institute for support during the course of this work.

\end{document}